\documentclass[12pts, draftcls, onecolumn]{IEEEtran}
\usepackage[latin1]{inputenc}
\usepackage[english]{babel}
\usepackage[T1]{fontenc}
\usepackage{array}
\usepackage{graphicx}
\usepackage[caption=false,font=footnotesize]{subfig}
\usepackage{fixltx2e}
\usepackage{cite}
\usepackage{algorithm}
\usepackage{algorithmic}
\usepackage{xfrac}
\usepackage{wrapfig}    
\usepackage[cmex10]{amsmath}
\usepackage{amsfonts,amssymb}
\usepackage{booktabs}
\usepackage{color}
\usepackage{epstopdf}
\usepackage{float}
\usepackage[T1]{fontenc}
\usepackage[cmex10]{amsmath}
\usepackage{amsfonts,amssymb}
\usepackage{multicol}
\usepackage{url}
\usepackage{yfonts}

\usepackage{enumerate}
\usepackage{tikz}
\definecolor{gold}{rgb}{0.85,.66,0}

\begin{document}
\title{Efficient Detectors for MIMO-OFDM Systems under Spatial Correlation Antenna Arrays}
\author{David William Marques Guerra, Rafael Masashi Fukuda, Ricardo Tadashi Kobayashi, and Taufik Abrão, \IEEEmembership{Senior, Member IEEE}
\thanks{Submission on May 5th, 2018. This work was supported in part by The National Council for Scientific and Technological Development (CNPq) of Brazil under Grants 304066/2015-0, and in part by CAPES - Coordenação de Aperfeiçoamento de Pessoal de Ní­vel Superior, Brazil (scholarship),  and by the Londrina State University - Paraná State Government (UEL).}
\thanks{D. W. Marques Guerra, R. M. Fukuda, R. T.  Kobayash and T. Abrão are with Universidade Estadual de Londrina (UEL), Londrina, Paraná, Brazil, taufik@uel.br, \,\,\, }
}
\maketitle

\begin{abstract}
This work {analyzes} the performance of the implementable detectors for multiple-input-multiple-output (MIMO) orthogonal frequency division multiplexing (OFDM) technique under specific and realistic operation system conditions, including antenna correlation {and array configuration}. Time-domain channel model has been used to evaluate the system performance under realistic communication channel and system scenarios, including different channel correlation, {modulation order}  and antenna arrays configurations. A bunch of MIMO-OFDM detectors were {analyzed} for the purpose of achieve high performance combined with high capacity systems and manageable computational complexity. Numerical Monte-Carlo {simulations} (MCS) demonstrate the channel selectivity effect, while the impact of the number of antennas, adoption of linear against heuristic-based detection schemes, and the spatial correlation effect under linear and planar antenna arrays are {analyzed} in {the} MIMO-OFDM context.
\end{abstract}


\IEEEpeerreviewmaketitle

\section{Introduction}
OFDM is a {modulation} scheme widely used in many communication systems, including {several} commercial applications such as wireless networks (Wi-Fi 802.11) and cellular systems (LTE) \cite{Hanzo2011}. In {those} systems, it is also {common to combine the} OFDM with {Multiple Input Multiple Output} (MIMO){, which can improve the spectral efficiency of the system \cite{Foschini1998}, \cite{Telatar1999}}. However, to {couple the OFDM to} the MIMO system, it is necessary to understand the basics of SISO channel and SISO OFDM. 

Usually, {inside an} OFDM system, a large number of subcarriers $N$ is deployed  in order to achieve a flat fading condition on each subchannel. This is particularly important in realistic scenarios, {where wireless} channel {introduces} dispersion effects on the signal, creating selective channels. In \cite{Saeed2003}, a SISO OFDM system was simulated to show how the number of subcarriers influence its performance on a multipath fading indoor channel based on Saleh-Valenzuela model, but not considering the Doppler frequency.

In flat fading channels, the coherence bandwidth of the channel $(\Delta B)_\textsc{c}$ is larger than the bandwidth of the signal, $W$. Hence, all frequency components of the signal will experience the same magnitude of fading. On the other hand, in frequency-selective fading channels $(\Delta B)_\textsc{c}< W$ occurs. As a consequence, different frequency components of the signal experience correlated fading.

{{In order to} verify the influence of the number of {subcarriers} $N$ on the system performance, the system must operate in multipath channels}. In the literature, some channel models are described in time domain (TD), {\it e.g.} Jakes model \cite{Proakis2006.ICI}, while other in frequency domain (FD), such as Clarke-Gans model \cite{Vanderlei04}.  In this article, the Jakes modified model proposed in \cite{Dent1993} is deployed. This TD channel simulator model performs a sum of cosines with random phases, multiplied by Walsh-Hadamard coefficients such that each waveform becomes uncorrelated.  In this model, the frequency Doppler is considered, {\it i.e.}, the mobility of the user terminal (UT) is taken into account aiming {to analyze} more realistic mobile radio scenarios.

In OFDM systems, to mitigate the inter-symbol interference (ISI) caused by multipath fading, it is necessary the use of guard interval. The most used type of guard interval on OFDM {systems} is the cyclic prefix (CP), as described analytically in \cite{Steendam_1999}.

One of the most recent well-established data transmission structure is multiple-input multiple-output (MIMO) system, which use multiple antennas {at the transmitter and receiver} sides to transfer data over a wire or wireless channels. MIMO systems are able to increase data rates by means of multiplexing or to improve performance/reliability through diversity mode \cite{Goldsmith2005}. The data increase can be achieved sending different data via different antennas. Sending {simultaneously} the same data via multiple antennas {the} {reliability} {are increased} exploiting diversities, such as time and space diversity. In spatial multiplexing, the signal that reaches {at} each receive antenna is interfered by the others $N_t - 1$ antennas, where $N_t$ represents the number of transmitting antennas. So, the purpose of demultiplexing-detection schemes is to mitigate the effects of the interference \cite{Hampton:2014}. Hence, on the receive side, there is a large number of MIMO detection techniques available. In this work, a bunch of MIMO-OFDM detectors are {characterized} and numerically {evaluated} under specific but realistic channel and system scenarios, including: the maximum likelihood (ML), the linear zero forcing (ZF) and the linear minimum mean square error (MMSE) detectors. Moreover, {two MIMO-OFDM detectors based on the evolutionary heuristic approaches also have been {analyzed}, namely the {\it particle swarm optimization} (PSO) detector, and {\it {differential} evolution} (DE) detector.}

Indeed, since the ML detector solution requires an exhaustive search going through all possible symbols combinations \cite{trimeche2013}, while linear closed solutions such as ZF and MMSE result in {a} poor performance for highly correlated channels \cite{Guerra_2016}, evolutionary heuristic algorithms are strong candidates to produce better solutions compared with linear detectors, and they results in reduced computational complexity compared to ML, since heuristic approaches do not evaluate all possibilities.

The PSO algorithm was already applied to solve the detection problem in MIMO-OFDM systems in \cite{trimeche2013,Seyman_2014}. In \cite{trimeche2013}, the PSO {and in \cite{Khan2007} the binary PSO (BPSO) are} evaluated and {numerical results of BER and computational} complexity {are analyzed}. In \cite{Seyman_2014}, the performances of DE, PSO and genetic algorithm (GA) are compared. On the other hand, in our work, {the performance-complexity tradeoff of the evolutionary heuristic PSO and DE MIMO-OFDM detectors are analyzed} under practical and useful scenarios, {\it i.e.},  considering {spatial correlated channels and other linear conventional MIMO-OFDM detectors}. The system model in the real-valued representation is considered while the selection procedure for the heuristic input parameters of PSO and DE algorithms are addressed accordingly. Besides, to the best of our knowledge, there are no works considering a comparative analysis of {evolutionary heuristics} and classical MIMO-OFDM detectors operating under spatial correlation antenna arrays.

{The {\it contribution} of this work is threefold. {\bf a}) Firstly, we analyze and compare a bunch of MIMO-OFDM detectors, including linear and evolutionary heuristic approaches, operating under realistic system configurations, regarding performance and implementability.  {\bf b})  Second, the influence of parameters related to the distance between the antennas, which determine the spatial antenna correlation, is discussed; two antenna arrays configurations are considered, the {\it uniform linear array} (ULA) \cite{Zelst2002} and the {\it uniform rectangular array} (URA) \cite{Levin2010}.  {\bf c}) Last, a {systematic} procedure is developed {and used} to calibrate the input parameters of both evolutionary heuristic PSO and DE detectors aiming at establishing a fair performance comparison between the linear and heuristic MIMO-OFDM detectors.}

The rest of this work is {organized} as follows. In section \ref{sec:ofdm}, the OFDM system is revised, {and time-domain channel emulator is explored, including details of modified Jakes channel model. Descriptions for the spatial channel correlation, the ML, ZF, MMSE, as well as the evolutionary heuristic PSO and DE detectors are developed in section \ref{sec:mimo}. Extensive numerical simulation results are analyzed in section \ref{sec:simulation_results}, including reliability evaluation, channel selectivity effect,  bit error rate (BER) performance comparison regarding spatial correlation, modulation order and sensibility analysis as well. Conclusions and final remarks are offered in \ref{sec:conclusions}.}

\noindent{\it Notation}: $\mathcal{F}$ and $\mathcal{F}^{-1}$ represent, respectively, the Fourier transform and the inverse Fourier transform; $\ast$ represents convolution operator. $[.]^H$ represents Hermitian operator. $\|.\|$ means Frobenius norm, $\mathbb{E}\{.\}$ expectation operator; bold lowercase letter represents a vector, while bold capital letter represents matrix. $\mathfrak{R\{.\}}$ and $\mathfrak{I\{.\} }$ operators represent the real and imaginary parts of a complex number. Operator $\circ$ denotes Hadamard product, while $\otimes$ is the Kronecker product.

\section{OFDM Transmission {and MIMO Channel}}\label{sec:ofdm}
OFDM is one type of multicarrier modulation that can be easily implemented using discrete Fourier transform (DFT) and its inverse (IDFT), or their equivalents fast Fourier transform (FFT) and inverse fast Fourier transform (IFFT). OFDM modulation consists in parallel data transmission with some modulation as M-QAM, M-PSK, etc, applying an IFFT to pass the signal of frequency domain to time domain. Thereafter, the cyclic prefix is added. Data is converted to an analogical signal. And finally, the signal is multiplied to a carrier with frequency {$f_c$} to transmit. 

On the receiver side, the signal $r(t)$ represents the transmitted signal $s(t)$ corrupted by noise. The signal $r(t)$ is multiplied by $\cos(2\pi f_c t)$, pass through a low-pass filter (LPF), then, the signal is converted to digital information, the cyclic prefix is removed and the serial data is converted to parallel. The DFT is performed, the symbols are converted to serial and demodulated to its respective scheme of modulation and the information bits can be estimated.

In order to mitigate intersymbol interference (ISI), some strategies, such as cyclic suffix, silence or the most common cyclic prefix (CP) can be adopted. 
CP consists in copying the last $\mu$ elements of the input sequence $s[n]$ and adding them to the start of $s[n]$, where {$h[n] = h[0], h[1], \dots h[\mu]$} represents the channel impulse response, with length $\mu+1$. After the CP addition, OFDM symbol becomes $\tilde{s}[n]$, with length $[N+\mu]$. Observe that the CP is an overhead, not carrying any information, which reduces the spectral efficiency.

The choice of the number of subcarriers ($N$) depends on the channel characteristics. For the design of an OFDM system, it is considered two properties of the channel, which are the maximum delay spread ($\tau_{\max}$) and maximum Doppler frequency $(f_{\textsc{d}})$. OFDM systems require that $N$ must be large enough so each subcarrier experiences a flat fading condition. Each subcarrier has a bandwidth $B$ smaller than the system total bandwidth, centered at a frequency $\omega_1, \omega_2, \dots, \omega_n$. Subcarriers bandwidth of $B$ can be overlapped at a maximum rate of 50\%.
\vspace{-3mm}

\subsection{MIMO-OFDM System}
\begin{figure}[!htb]
\centering
\subfloat[{Transmitter}]{%
\includegraphics[width=.45\textwidth]{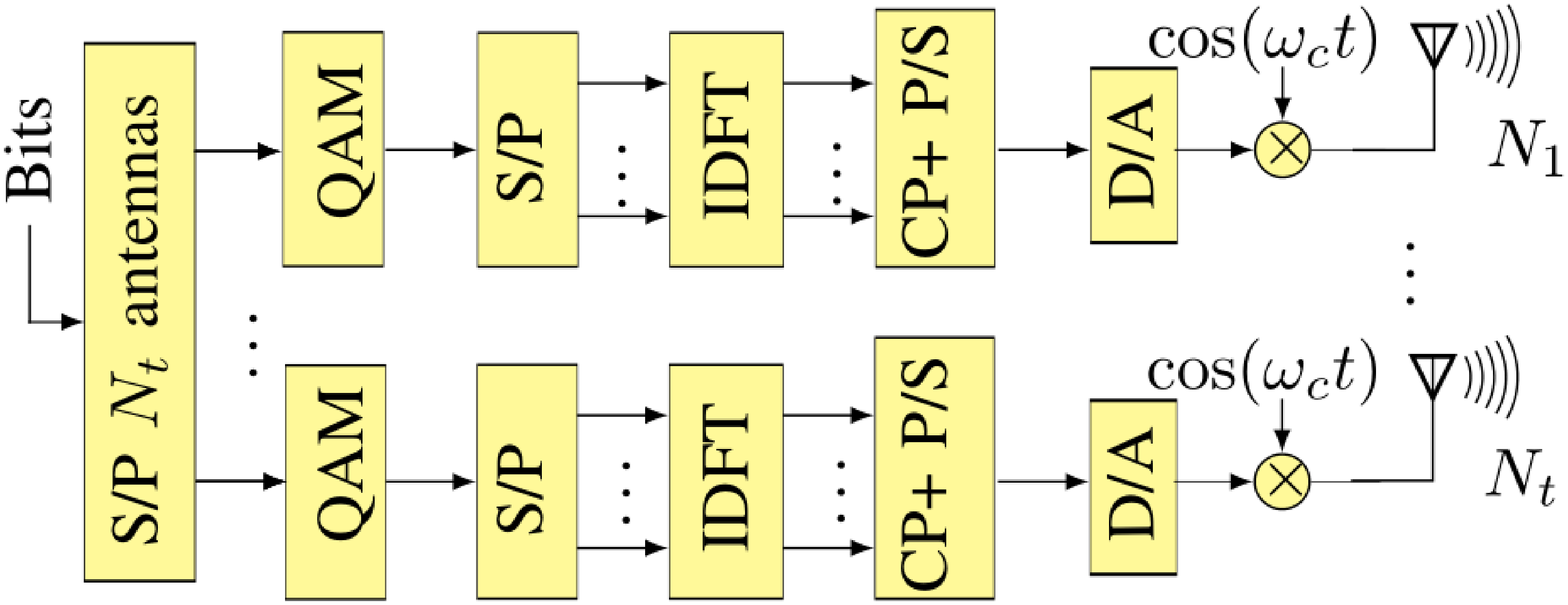}}
\label{a}\hfill
\subfloat[{Receiver}]{%
\includegraphics[width=.45\textwidth]{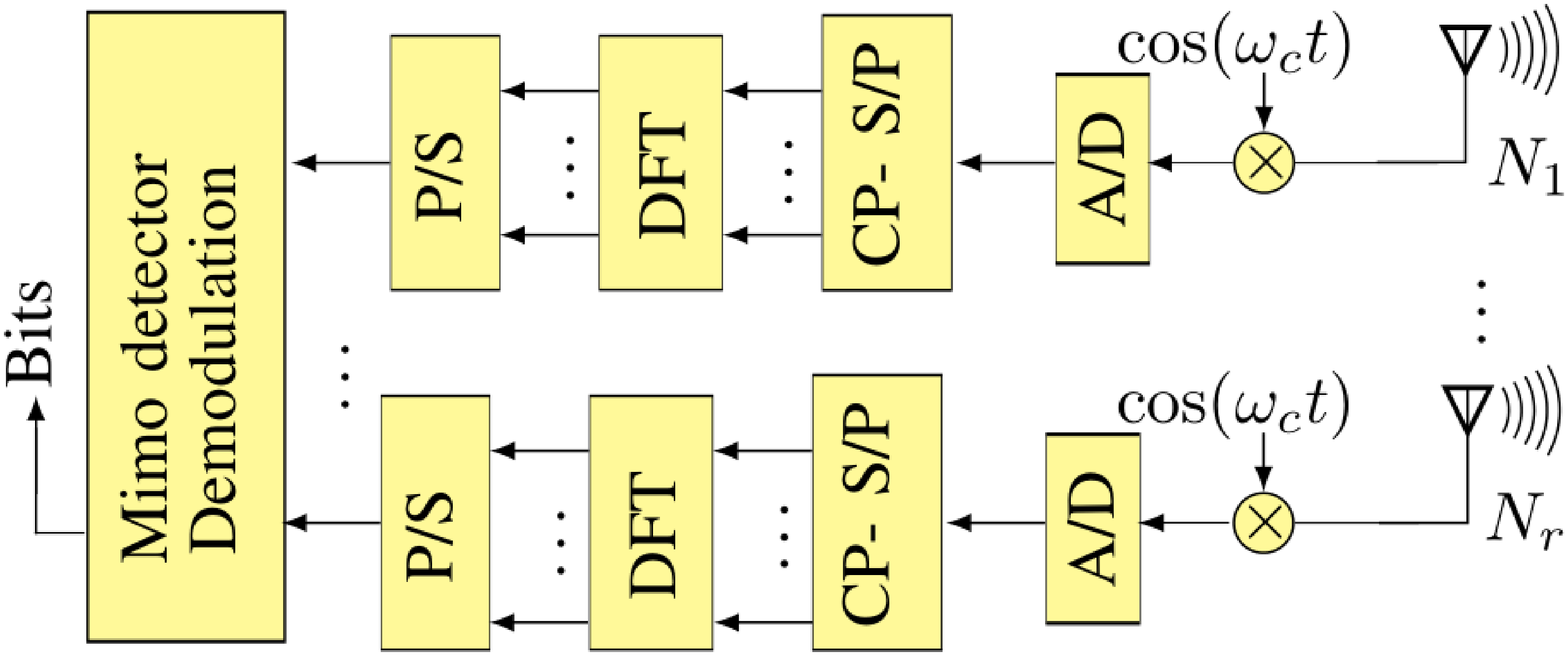}}
\label{b}\\
\caption{Block diagram of a MIMO-OFDM system.} 
\label{fig:mimo_ofdm_block_diagram}
\end{figure}

The combination of the OFDM system with the use multiple antennas at the transmitter and receiver results in MIMO-OFDM system, Fig. \ref{fig:mimo_ofdm_block_diagram}, {with $N_t$ transmit and $N_r$ received antennas}. QAM modulator and multiplexing configuration, {\it i.e.}, different data is sent through different antennas resulting in higher data rates than single-input-single-output (SISO) channel configuration have been considered.

On the transmitter side, {the data feeds a serial-to-parallel converter,} resulting in $N_t$ data streams, that are modulated in a similar way as OFDM SISO: the bit stream is modulated, the symbols are converted to parallel, the IDFT is performed, the cyclic prefix is added, the signal is multiplied by the carrier with frequency $f_c$ and finally transmitted. On the receiver side, the signal is converted to baseband, transformed to digital, the cyclic prefix is removed, and the signals serve a MIMO detector, and finally, the symbols are demodulated deploying QAM demodulator.

{As the OFDM technique allows parallel transmission over several subchannels, we can model a MIMO-OFDM system with $N$ subcarriers in time domain as \cite{Janhunen2011}, \cite{Paulraj2003}:}
\begin{equation}\label{eq:mimo-ofdm}
{\mathbf{y}[n] = \mathbf{H}[{n}]\mathbf{x}[{n}] +\mathbf{{z}}[n],\quad  \,\,\,\,\,\, {n} = 0,\,1,\,\dots,\,N-1}
\end{equation}
{where $n$ is the subcarrier index; $\mathbf{y}[n] \in \mathbb{C}^{N_r\times 1}$ is the received signals,  $\mathbf{H}[n] \in \mathbb{C}^{N_r\times N_t}$ is the channel matrix gains; $\mathbf{x}[{n}] \in \mathbb{C}^{N_t{\times 1}}$ is the transmit symbols and $\mathbf{{z}}[{n}] \in \mathbb{C}^{N_r{\times 1}}$ is the Gaussian noise with zero mean and variance $\sigma_{{z}}^2$.}

\begin{figure}[htbp!]
\centering
\includegraphics[width=0.49\textwidth]{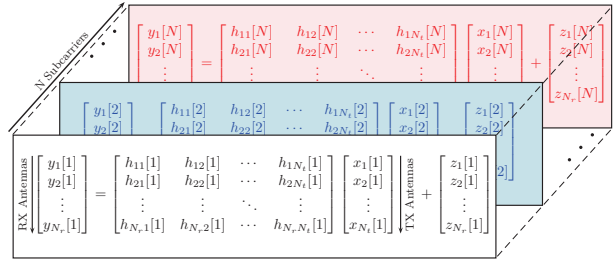}
\caption{{MIMO-OFDM problem.}}
\label{fig:MIMO_OFDM_f}
\end{figure}

{Therefore, we can interpret a MIMO system for each subcarrier, as illustrated in Fig. \ref{fig:MIMO_OFDM_f}. Thus, a MIMO-OFDM symbol block is composed of $N_r\times N_t$ OFDM symbols. Finally, it is important to note that if the number of subcarriers is insufficient to make the channel of each subcarrier flat, channel equalization can not be implemented correctly.}

Implementable MIMO-OFDM detectors operating in realistic fading channels and practical system configuration are discussed in Section \ref{sec:mimo}.

\subsection{Wireless Channel}\label{sec:wireless}
The noise is represented in time domain as it taken in account after the IDFT operation. In TD, fading interference can be written as a convolution operation of the channel impulse response $h(t)$ with the transmitter's signal $s(t)$,  which can be represented with discrete sequences:
\begin{equation}
\arraycolsep = 1.4pt
\begin{array}{rcl}
r[n] & = & s[n]\ast h[n].
\end{array}
\label{eq:circ_conv}
\end{equation}

The original information can be recovered analyzing the operation in FD, using the Fourier transform convolution property:
\begin{eqnarray}\label{eq:freq_eq}
s[n] = \mathcal{F}^{-1} \bigg\{\dfrac{\mathcal{F}\{r[n]\}}{\mathcal{F}\{h[n] \}}\bigg\}.  
\end{eqnarray}

Two important parameters in a wireless channel are the coherence time $(\Delta t)_c$ and the coherence band $(\Delta B)_\textsc{c}$. They are related to maximum Doppler frequency ($f_{\textsc{d}}$) and root-mean-square delay spread ($\tau_\textsc{rms}$), respectively. 
\begin{eqnarray}\label{eq:deltabc}
(\Delta t)_\textsc{c} = \dfrac{1}{f_{\textsc{d}}},\qquad (\Delta B)_\textsc{c} = \dfrac{1}{2\pi\tau_\textsc{rms}}.
\end{eqnarray}

The number of subcarriers in an OFDM system is defined by the channel's characteristics given in terms of delay spread, which can be observed experimentally via power delay profile (PDP), a graph that represents the channel parameters of power versus delay.

The PDP used in this work is the IEEE 802.11b, characterized by a decreasing exponential behaviour. The implementation of this channel model is discussed in several works in the literature, e.g., \cite{Cho2010}.

\subsection{Modified Jakes Channel Model} \label{sec:jakes_modified_model} 
In \cite{Jakes_mobile}, Jakes discusses statistical properties of wireless channels. The mathematical model described in time domain, namely Jakes model, performs a sum of cosines with angles uniformly distributed and considers mobility of Tx-Rx via the maximum Doppler shift effect. 
Twenty years later, Dent {\it et al} \cite{Dent1993} proposed improvements on the Jakes model, known as Jakes modified (JM) model. In this model, a factor subtraction of $0.5$ is suggested to eliminate the singular angles $0$ and $\pi$ [rad], while the multiplication of the fading samples by Walsh-Hadamard coefficients is indicated to generate uncorrelated channel waveforms. Herein, both the amplitude and phase of fading samples are analyzed and the JM model is implemented and applied to the analysis of OFDM systems.

To generate $k$ uncorrelated channel waveforms with Jakes modified model, the following relation must be respected:
\begin{equation}
C(t, k)=\sqrt{\dfrac{2}{N_{d}}}\sum_{n=1}^{N_d}{H_k(n)e^{j\phi_n}\cos{(\omega_n t + \theta_n)}},
\label{eq:C_coeff}
\end{equation}
where $H_k(n)$ represents the $k$-th Walsh-Hadamard (WH) coefficient, generated via WH matrices \cite{horadam2007}, $N_{d}$ is the number of oscillators, $\theta_n \sim \mathcal{U}[0, 2\pi]$ and $N_0=4\cdot N_d$, 
\begin{eqnarray}
\phi_n = \dfrac{\pi n}{N_d}, n = 1,2,\dots,N_d,\\
\alpha_n = \dfrac{2\pi (n-0,5)}{N_0},\\
\omega_n = \omega_m\cdot \cos{(\alpha_n)},\\
\omega_m = 2 \pi f_{\textsc{d}},
\end{eqnarray}
The autocorrelation function $\Phi(\tau)$ of the fading channel samples generated via JM method follows the Bessel function of first kind and 0th order:
\begin{equation}
\Phi(\tau) = J_0(\omega_m \tau).
\end{equation}
By applying the Fourier transform on the autocorrelation function, power spectrum density (PSD) can be derived:
\begin{equation}
\textsc{psd}(f) = \mathcal{F}\{J_0(\omega_m \tau)\} =  \frac{1}{\pi f_\textsc{d} \sqrt{1 - \left(\frac{f}{f_\textsc{d}}\right)^2}}.
\end{equation}

\section{{MIMO Spatial Correlation and Linear Detectors}}\label{sec:mimo}

\subsection{{MIMO-OFDM Spatial Correlation Model}}
In channel modelling, the correlation among transmit and/or receive antennas is an important aspect to be considered in realistic MIMO channels and systems \cite{Cho2010}. To model and evaluate the spatial antenna correlation, the Kronecker operator is deployed as:
\begin{eqnarray}
{{\mathbf{H}_{\rm corr}[n]} = \sqrt{{\bf R}_r}{\mathbf{G}[n]} \sqrt{{\bf R}_t^{H}},} 
\end{eqnarray}
{where ${\mathbf{H}_{\rm corr}[n]}$ is the correlated channel of the \textit{n}th subcarrier, uncorrelated channel matrix $\bf G$ is composed by independent and identically distributed (i.i.d.) entries, $\sqrt{{\bf R}_r}$ and $\sqrt{{\bf R}_t}$ are the square root of the spatial correlation matrices at the transmitter and receiver antennas, respectively.}

\subsection{Uniform Linear Antenna Array (ULA)}
A spatial correlation model for ULA is proposed in \cite{Zelst2002}.  This model considers that the antennas are arranged equidistantly, where $d_t$ and $d_r$ represent the spacing between the transmitting and receiving antennas, respectively. For simplicity of analysis, assuming the same number of antennas at the transmitter and receiver ($N_t = N_r$) side, while the spatial correlation matrix of the transmitter and receiver antennas are assumed equals ($\sqrt{{\bf R}_r}=\sqrt{{\bf R}_t}$). The spatial correlation matrix results Toeplitz, being expressed by:
\begin{eqnarray}
{\bf R}_t = {\bf R}_r =
\begin{bmatrix}
1     & \rho  & \rho^4 & \dots & \rho^{(N_t-1)^2} \\
\rho  & 1     &       &       & \vdots \\
\rho^4 & \rho  & 1     &       & \rho^4 \\
\vdots & \vdots & \vdots & \ddots & \rho \\
\rho^{(N_t-1)^2} & \dots & \rho^4 & \rho  & 1 \\
\end{bmatrix}
,
\end{eqnarray}
where $\rho\in [0,\,1]$ represents the normalised correlation index between the antennas.

\subsection{Uniform Rectangular Antenna Array (URA)}
An aproximation for the URA correlation model is proposed in \cite{Levin2010}. This model gives that the URA matrix correlation between the antennas is obtained from the Kronecker product of two ULA correlation matrices. Considering an URA configuration on the \textit{XY} plane, with $n_x$ and $n_y$ antenna elements along \textit{X} and \textit{Y} coordinates, respectively, so we have an array with $n = n_r \times n_y$ antennas. Considering that the correlation between the elements along \textit{X} coordinate does not depend on \textit{Y} and is given by matrix $\mathbf{R}_x$, and the correlation along Y coordinate does not depend on \textit{X} and is given by matrix $\mathbf{R}_y$. As a result, the Kronecker model approximation for the URA correlation matrix follows:
\begin{eqnarray}
\mathbf{R}_r = \mathbf{R}_x \otimes \mathbf{R}_y
\label{uracorr}
\end{eqnarray}

\subsection{Maximum likelihood (ML) MIMO Detector}
The ML detector provides the best performance between the detectors, but its complexity makes it impractical for real applications. This detector calculates all the possible symbols combinations and choose the one symbol vector {$\bf x$} which provides the minimum Euclidian distance between the received data {$\bf y$} and the reconstructed one defined by the channel matrix $\bf H$ and the symbol-vector candidate {$\bf x$}. Hence, the estimated symbol {$\boldsymbol{\mathrm{\tilde{x}}} $} can be mathematically expressed by:
\begin{eqnarray}\label{eq:MLD}
{\tilde{\bf x} = \min_{\bf x} \| \boldsymbol{\mathrm{y}} - \boldsymbol{\mathrm{Hx}}\|^2.}
\end{eqnarray}

\subsection{Zero-Forcing (ZF) MIMO Detector}
Considering a MIMO system operating under multiplexing mode, the data that reaches the receptor is the linear superposition of the signals of all the $N_t$ antennas \cite{Hampton:2014}. The ZF detector ignores the additive noise {\textbf{z} in eq. \eqref{eq:mimo-ofdm}} and solves the linear system multiplying the received signal by the inverse matrix which is defined, according to Moore-Penrose inverse matrix, as:
\begin{eqnarray}\label{eq: ZFD}
{\bf H}^\dagger_{\rm zf} = ({\bf H}^H {\bf H})^{-1} {\bf H}^H.
\end{eqnarray}
The estimated symbol is given by:
\begin{eqnarray}
{\tilde{\bf x} = {\bf H}^\dagger_{\rm zf} \,{\bf y}.}
\end{eqnarray}

\subsection{Minimum Mean Square Error (MMSE) MIMO Detector}
The MMSE detector considers the thermal noise channel statistics. This method tries to minimize the squared error between the true and estimated values of the transmitted symbols, {$\bf x$} and {$\tilde{\bf x}$}, respectively \cite{Hampton:2014} via optimisation
\begin{eqnarray}
{{\bf H}^\dagger_{\rm mmse}  =  \min_{\bf W}\,\,  \mathbb{E}\left\{\|\boldsymbol{\mathrm{y}} - \boldsymbol{\mathrm{Wx}}\|^2\right\}.}
\end{eqnarray}
Hence, solving this MMSE optimisation problem, the MIMO channel matrix results  in the MMSE pseudo-inverse matrix described as:
\begin{eqnarray}\label{eq:MMSED}
{\bf H}^\dagger_{\rm mmse} = \left({\bf H}^H {\bf H} + \dfrac{N_0}{E_S}{\bf I} \right)^{-1}{\bf H}^H.
\end{eqnarray}
where $\frac{N_0}{E_S}$ is the inverse of the signal-to-noise ratio (SNR).
Finally, the estimated symbol under linear MMSE MIMO detection is obtained in the same way of \eqref{eq: ZFD}, and given by:
\begin{eqnarray}
{\tilde{\bf x} = {\bf H}^\dagger_{\rm mmse} \,{\bf y}.}
\end{eqnarray}

\section{ {Heuristic-based MIMO-OFDM Detectors} }
{In this section, the heuristic PSO and DE algorithms are described in the context of the MIMO-OFDM detection problem. The complex system model is described in a well-known equivalent real-valued representation, {\it e.g.} \cite{proakis2008}. The deployment of the fitness function to evaluate the candidate-solution provided by heuristic algorithms is illustrated. PSO and DE algorithms are presented in the sequel, while the input parameters tuning problem for the evolutionary heuristic algorithms is addressed.}

\subsubsection{{ Real Value Representation} }
{The MIMO-OFDM system presented in eq. \eqref{eq:mimo-ofdm} can be represented as real-valued matrix and vectors in the form:}
{
\begin{equation}\label{eq:mimo-ofdm-real}
	\boldsymbol{\textswab{y}}[n]=\boldsymbol{\mathcal{H}}[n]\boldsymbol{\textswab{x}}[n] + \boldsymbol{\textswab{z}}[n],
\end{equation}
}
{with:}
\begin{equation}\nonumber
\boldsymbol{ \mathcal{H}}{[n]} = 
\begin{bmatrix}
\mathfrak{R}\{ {\bf H}{[n]} \}     &  -\mathfrak{I}\{ {\bf H}{[n]} \}     \\
\mathfrak{I}\{ {\bf H}{[n]} \}     &  \mathfrak{R}\{ {\bf H}{[n]} \}
\end{bmatrix}
, \quad \boldsymbol{\textswab{y}}{[n]} = 
\begin{bmatrix}
\mathfrak{R}\{ {\bf y}{[n]} \}     \\  \mathfrak{I}\{ {\bf y}{[n]} \}
\end{bmatrix}
, 
\end{equation}
\begin{equation}\nonumber
\qquad \boldsymbol{\textswab{x}}{[n]} = 
\begin{bmatrix}
\mathfrak{R}\{ {\bf x}{[n]} \}     \\  \mathfrak{I}\{ {\bf x}{[n]} \}
\end{bmatrix}
, \qquad \boldsymbol{\textswab{ {z} }}{[n]} = 
\begin{bmatrix}
\mathfrak{R}\{ {\bf z}{[n]} \}     \\  \mathfrak{I}\{ {\bf z}{[n]} \}
\end{bmatrix}
,
\end{equation}
where $\boldsymbol{\mathcal{H}}{[n]} \in \mathbb{R}^{2N_r \times 2N_t}$ is the real-valued representation of the channel matrix, vectors $\boldsymbol{\textswab{x}}{[n],\textswab{ {z} }}{[n]} \in \mathbb{R}^{2N_t \times 1} $ are the real-valued representations of the original information and additive noise, respectively, and $\textswab{y}{[n]} \in \mathbb{R}^{2N_r \times 1}$ the real-valued received signal.

\subsubsection{{ Fitness Function} }
The fitness function evaluates {the quality of the estimated symbol  and guides the evolutionary heuristic search on the candidate-solution feaseable subspace. For detection problem, the fitness function} is based on the Euclidean distance between the received signal and the reconstructed one \cite{Seyman_2014,trimeche2013,Khan2007}. Considering the $k$th candidate-solution of a evolutionary heuristic, $\boldsymbol{\zeta}_k$, namely particle in PSO or individual in DE, the fitness function is calculated as:
\begin{equation}\label{eq:fitness}
f( \boldsymbol{\zeta}_k ) = \|\boldsymbol{\textswab{y}}{[n]}-\boldsymbol{\mathcal{H}}{[n]} {\boldsymbol{\zeta}_k} \|^2, 
\end{equation}
For the detection problem, a minimization problem is considered and lower values of fitness function are desired. 

\subsection{{PSO-based Detection Algorithm} }
PSO was proposed by \cite{Kennedy_1995} considering a population-based approach, emulating bird flocking and fish schooling {behavior}. The PSO algorithm performs calculation of velocity and position of each particle inside the swarm; {using matrix representation \cite{Cheng2011}, they are given respectively  by:}
\begin{equation}\label{eq:psoVelocity}
{\bf V} = w{\bf V} + c_1 {\bf U}_1 \circ ({\bf M}_{\textsc{pb}}{\bf - P}) + c_2 {\bf U}_2 \circ ( {\bf M}_{\textsc{gb}}{\bf - P} ),
\end{equation}
\begin{equation}\label{eq:psoPostion}
\text{and} \qquad \bf P = P + V,
\end{equation}
where $w, c_1, c_2$ {represent} inertia, cognitive and social factors, respectively; ${\bf U}_1$ and ${\bf U}_2$ are random matrices with elements following uniform distributions ${\bf U}_i\sim\mathcal{U} [0; 1]$; ${\bf M}_{\textsc{pb}}$ is a matrix that store values of personal best of each particle  and ${\bf M}_{\textsc{gb}}$ is a matrix constructed with the positions of the global best particle ${\bf p}_{\textsc{gb}}$, given in the form ${\bf M}_{\textsc{gb}} = [{\bf p}_{\textsc{gb}} \dots {\bf p}_{\textsc{gb}}] \in \mathbb{R}^{N_{\mathrm{dim}} \times N_{\mathrm{pop}} }$. ${\bf P}$ is a real-valued matrix representing positions, while ${\bf V}$ represents particle velocity matrix; explicitly: 
\begin{equation}\nonumber
{\bf P} = 
\begin{bmatrix}
{\bf p}_1  \dots {\bf p}_{N_{\textrm{pop}}}
\end{bmatrix}
, \quad
{\bf V} = 
\begin{bmatrix}
{\bf v}_1  \dots {\bf v}_{N_{\textrm{pop}}} 
\end{bmatrix}
\in \mathbb{R}^{N_\textrm{dim} \times N_\textrm{pop} }
\end{equation}
where vectors ${\bf p}_k, {\bf v}_k \in \mathbb{R}^{N_{\textrm{dim}} \times 1}$ with $k = 1, \dots, N_{\textrm{pop}}$ represent the position and velocity of {the \textit{k}th} particle, with $N_{\textrm{pop}}$ representing the population size and $N_{dim}$ the dimensionality of the problem.

{In order to avoid a possibly grow to infinity of the velocity vector \cite{Mohammad_2017}, the limitation of velocity $[-V_{\rm max}, V_{\rm max}]$ \cite{Shi1998b} was considered, where $V_{\rm max}$ represents the maximum achievable velocity of the $N_{\rm pop}$ particles. Regard the inertia parameter, it can be a constant, a linear or nonlinear function \cite{Shi1998a}. In this work, in order to give to the algorithm exploitation ability at the beginning and exploration to fine search the solution \cite{Shi1998b}, a decreasing strategy at each iteration of the inertia factor given by $0.99w$ is considered. }

{The initialization of the both implemented PSO and DE heuristic algorithms was the same; the position of the particles ${\bf P}$ and initial population in DE are generated randomly following an uniform distribution inside the search space of the problem \cite{Storn1997}. Those positions are set as the personal best position of the particle in the matrix ${\bf M}_{\textsc{pb}}$. The fitness function in eq. \eqref{eq:fitness} is evaluated ($\boldsymbol{\zeta}_{k} = {\bf p}_k, k = 1,\dots, N_{\rm iter}$) and the position of the particle that produces the lowest value (since we are dealing with a minimization problem) is set as the global best position ${\bf p}_{\textsc{gb} }$, and the matrix ${\bf M}_{\textsc{gb}}$ is formed. }

{After evaluation of eq. \eqref{eq:psoVelocity} and eq.\eqref{eq:psoPostion}, matrices ${\bf M}_{\textsc{pb}}$ and ${\bf M}_{\textsc{gb}}$ are updated (if needed) and the process is repeated till the stop criteria is met. In our implementation, the stop criterion based on the pre-defined maximum number of evaluations $N_{\rm iter}$ is considered. Hence, after $N_{\rm iter}$ iterations, the output of the evolutionary heuristic algorithm is the vector of the best position ${\bf p}_{\textsc{gb}}$, which is the estimated symbol $\tilde{\bf x}$ in the MIMO-OFDM detection problem. }

A pseudocode {summarizing the procedure} for the evolutionary heuristic PSO algorithm is presented in Algorithm \ref{algo:PSO}.
\begin{algorithm}[h]
\small
\caption{ PSO -- Particle Swarm Optimization.}\label{algo:PSO}
\begin{algorithmic}[1]
\STATE{ Input parameters: \, $c_1, c_2, w, N_{\mathrm{pop}}, N_{\mathrm{iter}}$}
\STATE{ {Generate initial positions ${\bf P}$ } }
\STATE{ {Fitness function evaluation and initialization of ${\bf M}_{\textsc{PB}}$, ${\bf M}_{\textsc{GB} }$ } }
\FOR {1 \TO  $N_{\mathrm{iter}}$}
\STATE{ Calculate velocity, eq. \eqref{eq:psoVelocity}}
\STATE{ Calculate position, eq. \eqref{eq:psoPostion}}
\STATE{ Evaluate fitness function, eq. \eqref{eq:fitness}, for all particles ${\bf p}_k$}
\STATE{ Update personal best matrix ${\bf M}_{\textsc{pb}}$}
\STATE{ Update global best matrix ${\bf M}_{\textsc{gb}}$}
\STATE{ {Velocity limitation} }
\STATE{ {Inertia factor reduction} }
\ENDFOR
\STATE{Output: \, ${\bf p}_{\textsc{gb}}$}
\end{algorithmic}
\end{algorithm}

\subsection{DE-based Detection Algorithm}
The differential evolution (DE) is an evolutionary population-based heuristic that relies on a population of individuals to find global optima. The algorithm relies on operations of mutation, crossover and selection to produce more suitable individuals through $N_{\rm gen}$ generations. 

The DE algorithm was presented in \cite{Storn1997} and operates as follows. There are $N_{\rm ind}\geq 4$ vectors of individuals that are represented as ${\boldsymbol{\iota}_{k}} \in \mathbb{R}^{N_{\rm dim}\times 1}, k = 1,\dots N_{\rm ind}$, where $N_{\rm dim}$ represents the dimensionality of the problem. Herein, following the procedure defined in \cite{Storn1997}, the \texttt{rand/1/bin} strategy is employed. {Strategies to scape to local optima adopted in DE-based detector are described in the following.}

\begin{enumerate}[a)]

\item {Mutation:}
The $k$-th mutation vector $\boldsymbol{\nu}$ is constructed as:
\begin{equation}\label{eq:mutation}
\boldsymbol{\nu}_{k} = \boldsymbol{\iota}_{r_1} + F_{\rm mut}(\boldsymbol{\iota}_{r_2} - \boldsymbol{\iota}_{r_3})
\end{equation}
where $k\neq r_1 \neq r_2 \neq r_3$, $k=1,\dots, N_{\rm ind}$. {Variables} $r_{{1}},r_{{2}},r_{{3}}$ are integer random {indexes} uniformly distributed {inside} the interval $[1,2,\dots, N_{\rm ind}]$ and $F_{\rm mut} \in [0, 2]$ represents the mutation scale factor. 

\item {Crossover: }
The $k$-th crossover vector $\boldsymbol{\psi}_{k}$ ($k=1,\dots,N_{\rm ind}$) is constructed as following. The $i$-th element, $i = 1,\dots, N_{\rm dim}$,  of the $k$-th crossover vector $\boldsymbol{\psi}_{k}$ is selected given the following rule:
\begin{equation}\label{eq:crossover}
\psi_{ik} = 
\begin{cases}
\nu_{ik}       & \quad \text{if } rand \in [0, 1] \leq F_{\rm cr} \text{ or } i = r_4\\
\iota_{ik}     & \quad \text{if } rand \in [0, 1] > F_{\rm cr} \text{ and } i \neq r_4 
\end{cases} 
\end{equation}
where $rand \sim \mathcal{U}{[0, 1]}$; $r_4$ is an integer randomly generated {in the interval} $[1, \dots, N_{\rm dim}]$; the crossover factor is defined by $F_{\rm cr} \in [0, 1]$. As pointed out in \cite{Storn1997}, the crossover vector has at least one element from the mutation vector, {\it i.e.}, the condition $i = r_4$. 

\item {Selection: }
The next generation of individuals $\boldsymbol{\iota}_{k}^{\textsc{g}}$ is constructed as:
\begin{equation}\label{eq:selection}
\boldsymbol{\iota}_{k}^{\textsc{g}} = 
\begin{cases}
\boldsymbol{\psi}_{k}       & \quad \text{if } f(\boldsymbol{\psi}_{{k}}) < f(\boldsymbol{\iota}_{{k}})\\
\boldsymbol{\iota}_{k}     & \quad \text{otherwise } 
\end{cases} 
\end{equation}
The fitness function {in eq.\eqref{eq:fitness}} evaluates $\boldsymbol{\iota}_k$ and $\boldsymbol{\psi}_k$. Vectors that produce more suitable values (smaller values) are selected and a new generation of individuals is produced.

\end{enumerate}

{After the execution of $N_{\rm gen}$ iterations, the best individual, in other words, the individual corresponding to the lowest value of fitness function in  eq. \eqref{eq:fitness} is the output of the algorithm, and the estimated symbol $\tilde{\bf x}$ of the MIMO-OFDM detection problem. A pseudocode} synthesizing DE steps is presented in Algorithm \ref{algo:DE}. 

\begin{algorithm}
\small
\caption{DE -- Differential Evolution.}\label{algo:DE}
\begin{algorithmic}[1]
\STATE{ Input parameters: \, $F_{\rm cr}, F_{\rm mut}, N_{\mathrm{ind}}, N_{\mathrm{gen}}$}
\STATE{Generate initial individuals}
\FOR {1 \TO  $N_{\mathrm{gen}}$}
\STATE{Mutation, eq. \eqref{eq:mutation}, $k = 1, \dots, N_{\rm ind}$}
\STATE{Crossover, eq. \eqref{eq:crossover}, $i=1,\dots,N_{\rm ind}; k = 1, \dots, N_{\rm ind}$ }
\STATE{ Select new individuals, eq. \eqref{eq:selection}, $k = 1, \dots, N_{\rm ind}$ }
           \ENDFOR
                \STATE{Output: \, best individual $\boldsymbol{\iota}$}
                \end{algorithmic}
                \end{algorithm}

\subsection{ {Input Parameters} }
The non-optimal input parameter values choice could substantially degrade the performance results provided by {the heuristic algorithm in a given} application, as studied in \cite{Marinello_2012} {for Ant-Colony Optimization algorithm. Besides, PSO algorithm also suffers alteration of convergence properties when input parameters are chosen incorrectly} \cite{Mohammad_2017,Clerc_2002,Trelea_2003}. {In a same way, the DE-based algorithm has some recommended interval of values to achieve fast convergence with DE algorithm \cite{Storn1997}.   For instance, the number of individuals must be in between $N_{\rm ind} \in \{5; \, 10\}N_{\rm dim}$, where $N_{\rm dim}$ is the problem dimension, as suggested in \cite{Storn1997}.}

{In order to provide a fair comparison between the selected evolutionary heuristic algorithms, and since the such approach} are sensible to the choice of the input parameter values, which can differ substantially considering different {optimization} problems nature, {the input parameter tuning procedure herein is obtained numerically, and discussed in subsections \ref{subsec:inputCalibPSO} and \ref{subsec:inputCalibDE}.}

\section{Numerical Results} \label{sec:simulation_results}
The OFDM scheme is numerically simulated at equivalent baseband frequencies and the associated {performance evaluated and analyzed} with respect to the number of subcarriers, modulation order and different SNR regions. For the simulation of OFDM system, a system with $4$-QAM, $16$-QAM and $256$-QAM modulation formats has been considered. The number of subcarriers was increased and the system performance in terms of BER {\it versus} $E_{b}/N_{0}$ analyzed. After the verification in terms of subcarriers number and prefix cyclic, numerical simulation results of MIMO-OFDM system are discussed. In this case, specifically linear and evolutionary heuristic detectors performance subject to spatial antenna correlation effect has been compared.

\subsection{Statistical Characteristics for Jakes Model Fading}
In this subsection samples generated by the implemented JM method is analyzed in order to verify whether or not their statistical properties are compatible with the theoretical channel model.

\begin{table}[htb]
\caption{Jakes modified simulation parameters. \label{tab:jakes_mod}}
\centering
\begin{tabular}{l l}    \hline
\textbf{Parameter}      & \textbf{Value} \\\hline\hline
                \# oscillators          & $N_d = 1024 $         \\
                noise samples                   &       16384   \\
                Doppler frequency &     $f_d = 83$ Hz   \\
                sampling period         & $T_s = 383.5 \mu$s\\
                $k$ waveforms                   &       4       \\
                \hline
                \end{tabular}
                \end{table}

\begin{figure}[!htb]
                \centering
                \includegraphics[width=.5\textwidth]{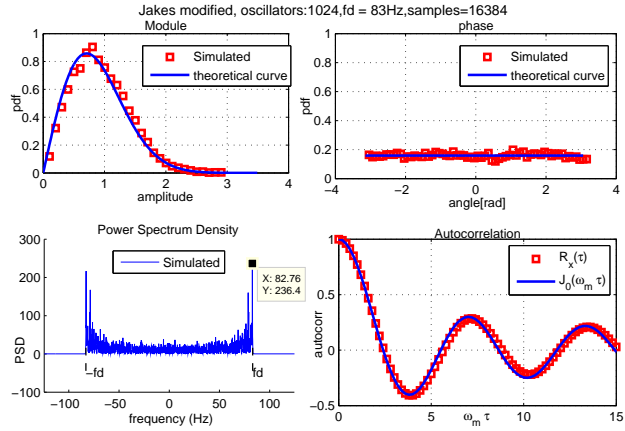}
 \vspace{-6mm}
 \caption{Fading samples: module, phase, PSD and autocorrelation.}
                \label{fig:jakes_mod}
                \end{figure}

The JM fading characteristics are presented in Fig. \ref{fig:jakes_mod}, which were generated using the JM parameters listed in Table \ref{tab:jakes_mod}. In these plots, the behaviour of only one  channel path waveform is depicted, since the statistical properties was similar for the other waveforms. As it can be observed, the module's probability density function followed a Rayleigh distribution, while the phase variable follows an uniform distribution with values between $-\pi$ to $\pi$. Also, The obtained autocorrelation function corroborates the theoretical Bessel 0-th order $J_0(\omega_m \tau)$ function; as expected, the PSD peaks were located at frequencies $f_c-f_\textsc{d}$ and $f_c+f_\textsc{d}$, with $f_c=0$ Hz, for convenience. 

\subsection{OFDM Performance under (non)-Selective JM Channel Generator}

\begin{figure}[htbp!]
\centering
\includegraphics[width=.48\textwidth]{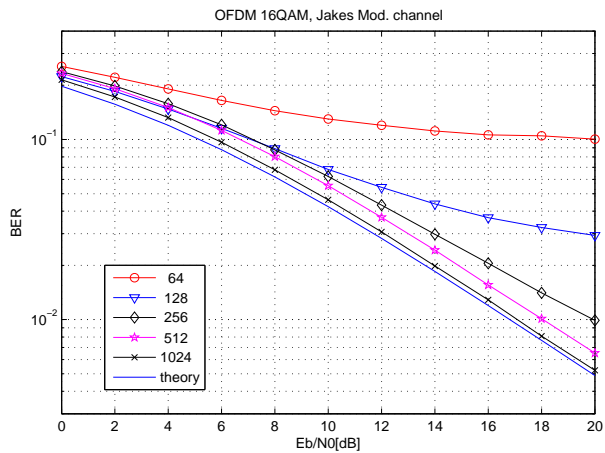}
\vspace{-4mm}
\caption{16-QAM OFDM system BER performance with fading coefficients generated by the modified Jakes model. The number of subcarriers were changed  to demonstrate the effect of the subchannel flatness on the OFDM performance.}
\label{fig:3ofdm_jakesmod}
\end{figure}

After observing that the generated samples are statistically consistent with the Jakes modified model and PDP described in \cite{Cho2010}, the BER performance varying the number of subcarriers was {analyzed}. It was simulated an OFDM system with 16-QAM and fading channel samples obtained from the JM model, where the number of subcarriers was varied from a selective sub-channel basis condition to a totally flat channel condition over each OFDM sub-channel. Curves for the BER performance are depicted in Fig. \ref{fig:3ofdm_jakesmod}.  The associated system parameters deployed in the simulation are summarized in Table \ref{tab:ofdm_pdp_jakes_mod3}. Indeed, we have considered a realistic scenario with OFDM system bandwidth $W = 5$ MHz, delay spread $\tau_\textsc{rms} = 2.5\mu$s and a low-median mobility, resulting in a maximum Doppler frequency of $f_\textsc{d} = 23$ Hz. Hence, the coherence time and the coherence bandwidth are readily determined by eq. \eqref{eq:deltabc}.

\begin{table}[htb]
\caption{OFDM simulation with PDP and Jakes modified model. \label{tab:ofdm_pdp_jakes_mod3} }
\centering
\begin{tabular}{l l}    \hline
\textbf{Parameter}      & \textbf{Value} \\\hline\hline
                \# oscillators          & $N_d = 256 $          \\
                Max. Doppler frequency &        $f_\textsc{d} = 23$ Hz  \\
                Jakes mod. sampling period              & $100\cdot f_{\textsc{d}}$\\
                System Bandwidth &  $W = 5$ MHz\\
                \# Subcarriers (N)      &       64, 128, 256, 512, 1024\\
                $\tau_{\textsc{rms} }$  &       2.5 $\mu$s\\ 
                PDP                             &       {IEEE 802.11b}\\
                \hline
                \end{tabular}
                \end{table}

                Since the fading coefficients are generated in time domain,  the circular convolution operation was performed between the coefficients in time domain (after IFFT operation) and JM fading coefficients. The fading effect was removed using eq. \eqref{eq:freq_eq}. 

                As it can be observed in Fig. \ref{fig:3ofdm_jakesmod} the BER performance of the system under such circumstances increases as the number of subcarriers grows. Also, the number of subcarriers for flatness condition of course depends on the coherence bandwidth of the channel. If the number is large enough, $N \gg \frac{W}{(\Delta B)_\textsc{c}}$, the flat channel condition is achieved.

                Deploying quadrature amplitude modulation of $M$-ary order ($M$-QAM), the BER performance of a OFDM system operating under fading channel tends to the $M$-QAM BER at Rayleigh fading channel. Analytical expression for $M$-QAM performance under fading channel is presented in \cite{Cho2010}.

Relating the coherence bandwidth with number of subcarriers, the flatness condition in OFDM subchannel is achieved if:
$$BW_{\rm ch} = \dfrac{W}{N} \ll (\Delta B)_c$$

For practical purpose, one can assume that the bandwidth of the OFDM subchannel is five times less than the coherence bandwidth:
$$ BW_{\rm ch} = \dfrac{W}{N} = 0.2 \cdot(\Delta B)_c$$

From eq. \eqref{eq:deltabc} one can define the minimum number of subcarrier for flatness condition:

$$N = \dfrac{W\cdot 2\pi\tau_\textsc{rms}}{0.2} \cong 393$$

\begin{figure}[htbp!]
\centering
\includegraphics[width=.48\textwidth]{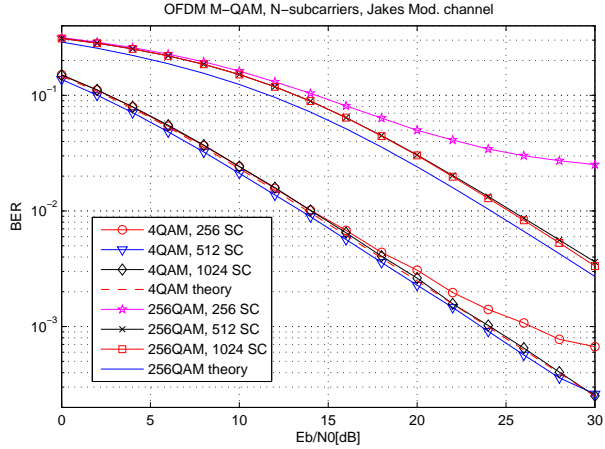}
\vspace{-4mm}
\caption{BER performance for OFDM $M$-QAM  with fading coefficients generated via JM  model. QAM order and number of subcarriers were changed.}
\label{fig:4ofdm_mqam}
\end{figure}

This condition is corroborated by the BER performance behaviour in Fig.\ref{fig:3ofdm_jakesmod} and Fig. \ref{fig:4ofdm_mqam}.

\vspace{-2mm}

\subsection{OFDM Performance under (non)-Selective JM Channel Model with different CP Sizes}
\begin{figure}[htbp!]
\centering
\includegraphics[width=.48\textwidth]{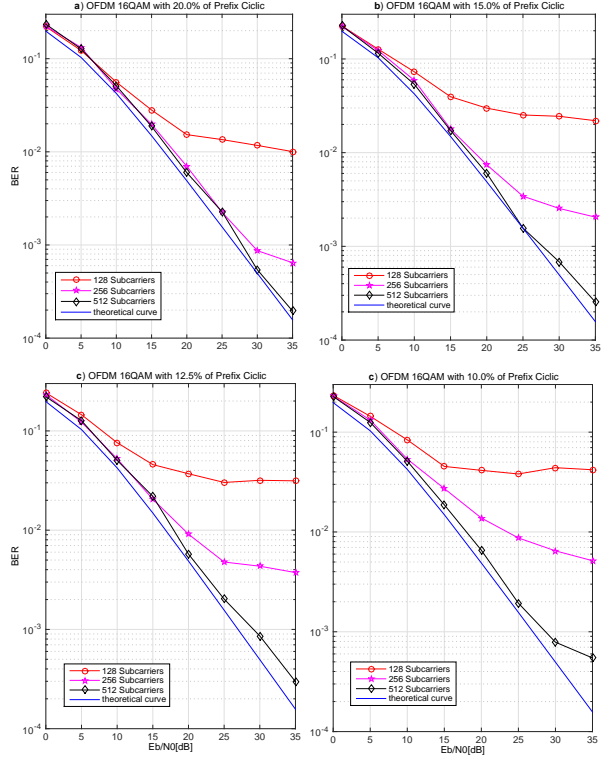}
\vspace{-4mm}
\caption{OFDM 16QAM performance with fading coefficients generated with JM model. CP size: a) $0.20\, T_\textsc{ofdm}$; b) $0.15 \, T_\textsc{ofdm}$; c) $0.125\, T_\textsc{ofdm}$; d) $0.10\, T_\textsc{ofdm}$.}\label{fig:ber_teste_cp}
\end{figure}

Considering the same scenario of the previous subsection with system bandwidth $W = 5$ MHz, delay spread $\tau_{\textsc{rms} } = 2.5\mu$s, the impact of cyclic prefix (CP) sizes on the OFDM performance is depicted in  Fig. \ref{fig:ber_teste_cp}.

As shown in Fig. \ref{fig:3ofdm_jakesmod}, the system achieve the flatness channel condition with $N=512$ subcarriers. When an excessive CP size is used (around 20\%), the performance of the system with 256 subcarriers approaches the performance of single-carrier flat fading condition. However, since the number of subcarrier $N$ is not enough for the OFDM system achieve the flatness condition, the system will present a limited performance, because selectivity of the channel in each subcarrier. Moreover the reduction in the CP size causes the OFDM inter-symbol interference (ISI), degrading the performance, i.e.,  even for an increase of SNR the BER performance remained virtually the same (BER floor effect).

Indeed, in flat channel condition ($N=512$ subcarriers), note that the system performance tends to deteriorate with decreasing of CP size. The system has about the same performance for all the considered values of CP, however, when the CP size is insufficient to combat ISI, which occurs for a CP around 10\% of OFDM period, the system starts to show a BER floor (limitation), whereas an increase in $E_b/N_0$ no longer provides a significant performance increase (decreased BER). However, if the CP size is very large we will be expending energy unnecessarily, because the CP increases the OFDM period to combat ISI and therefore waste energy to do this. From Fig. \ref{fig:ber_teste_cp},one should use a CP size from 12.5 \% to 15 \%, then the lowest energy expenditure possible with the CP can be achieved, in which performance remains very close to the theoretical.

\vspace{-2mm}
\subsection{MIMO-OFDM Reliability Evaluation}
In this section it was considered a similar scenario than the previous section, but observing that MIMO-OFDM system requires a higher computational and implementation complexity. The parameters adopted in Monte-Carlo simulations are shown in Table \ref{tab:mimo_OFDM}. Additionally, the system operates with perfect channel state information (CSI) and linear ZF and MMSE {equalizers}. Performance of such linear detectors is compared with the optimum maximum-likelihood (ML) MIMO-OFDM detector. The total power allocated is the same as with the SISO system, so it was equally distributed  (EPA) among the $N_t$ antennas in order to promote a fair comparison.

\begin{table}[!htb]
\caption{MIMO-OFDM simulation parameters. \label{tab:mimo_OFDM}}
\centering
\begin{tabular}{l l}    \hline
\textbf{Parameter}		&\textbf{Value} \\\hline\hline
\multicolumn{2}{c}{OFDM} \\
\hline
System Bandwidth, BW				&	20MHz     \\
Modulation order, $M$				&   4-QAM   	\\
Delay spread, $\tau_\textsc{rms}$ 	&	{51ns}   \\
\# Subcarriers, $N$      			&	64\\
{$(\Delta B)_c$}				&	{3.125MHz}\\
{Subcarrier flatness. $\frac{(\Delta B)_c}{BW/N}$}	 	&	{10}	\\
\hline
 \multicolumn{2}{c}{MIMO} \\
 \hline
\# Antennas, $N_t\times N_r$       & $2\times 2; \,\, 4\times4;$ \, \, $8\times8$\\
{Antenna array type} & {Linear (ULA); \,\, {Rectangular} (URA)}\\
Spatial correlation index & $\rho \in [0;\,\, 0.5; \,\, 0.9]$\\
                Linear detectors &  ZF \& MMSE\\
                Heuristic detector & PSO\\
                Power allocation strategy & EPA \\
                \hline
                \multicolumn{2}{c}{Channel} \\
                \hline
                Type & NLOS Rayleigh channel\\
                CSI knowledge & perfect\\
                Mobility (freq Doppler)& $f_{\rm D} = 0$ Hz\\ 
                \hline
                \multicolumn{2}{c}{ PSO Detector} \\
                \hline
                Population size $N_{\rm pop}$ & {40} \\
                Iterations $N_{\textrm{max}}$    & {100} \\
                Search Space    &    [-1; 1] \\
                Cognitive factor $c_1$   & 4 \\
                Social factor $c_2{(\rho)}$   & {1 ($0$) \,\,0.5 ($0.5$) \,\, 1 ($0.9$)} \\
                Inertia $w{(\rho)}$   &  {1.5 ($0$) \,\,1.5 ($0.5$) \,\, 3.5 ($0.9$)} \\
\hline
\multicolumn{2}{c}{{DE Detector}} \\
                \hline
         {    \# generation   $N_{\rm gen}$}  & 100 \\
                Crossover factor $F_{\rm cr}{(\rho)}$     & {0.6 ($0$) \,\,0.6 ($0.5$) \,\, 0.8 ($0.9$)} \\
                Mutation  factor $F_{\rm mut}(\rho){(\rho)}$ &  {0.6 ($0$) \,\,0.8 ($0.5$) \,\, 1.8 ($0.9$)} \\
                \# individuals $N_{\rm ind}$   & {40} \\
                \hline
                \end{tabular}
                \end{table}

Specifically, in the MIMO-OFDM detection problem with heuristics, a 4-QAM modulation format was considered, with valid symbols defined by $\{-1 + 1j, -1 - 1j, 1 + 1j, 1 - 1j \}$, while the search space was limited to the interval of integer values $[\pm 1]$. The heuristic algorithm was applied to each subcarrier as presented in the model description in eq. \eqref{eq:mimo-ofdm-real}, resulting in $N_{\rm dim} = 2N_t$ symbols to be estimated per subcarrier. Under PSO detection algorithm, the parameter $V_{\rm max}=1$, reflecting the dynamic range of each particle inside the search space \cite{Shi1998b} was considered in the simulations. 

\begin{figure}[htbp!]
                \centering
                \includegraphics[width=.48\textwidth]{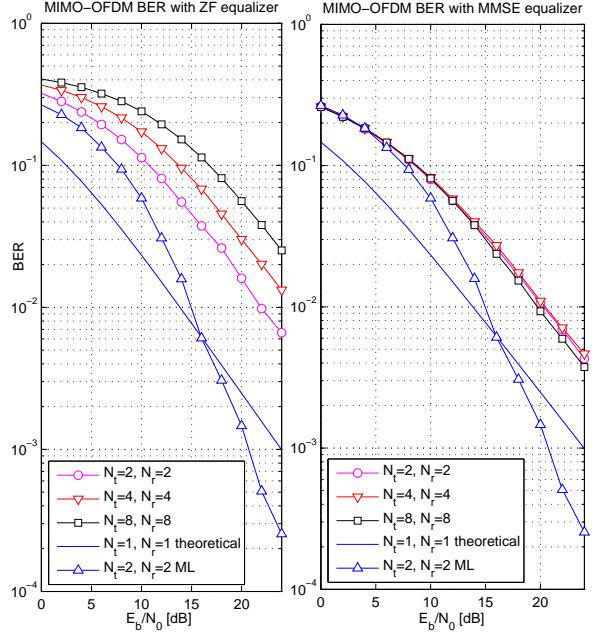}
                \vspace{-4mm}
                \caption{BER performance for MIMO-OFDM 4-QAM under flat fading channel, linear ZF and MMSE detectors, and different number of antennas.}
                \label{fig:mimo_ofdm_antennas}
                \end{figure}

The numerical results obtained changing the number of antennas from $2\times 2$ to $ 4 \times 4$ and $8 \times 8$ for ZF and MMSE equalization are presented in Fig. \ref{fig:mimo_ofdm_antennas}; one can see that as the number of antennas increases, BER performance MIMO-OFDM equipped with ZF detector deteriorates, while with MMSE detector, the BER performance does not have significant changes. As expected, the superior performance of MMSE detector over ZF detector can be explained since MMSE takes in account thermal noise statistics, while results more computationally complex than ZF. Analysing the numerical results in terms of data rate, the $8 \times 8$ system transmits 4 times more information than $2 \times 2$ configuration. With perfect knowledge of the channel and MMSE detector, $8 \times 8$ OFDM-MIMO system provides more data rates (multiplexing mode) with similar reliability (no substantial BER degradation) compared with the other antennas configurations.  

\begin{figure}[h!]
                \centering
                \includegraphics[width=.48\textwidth]{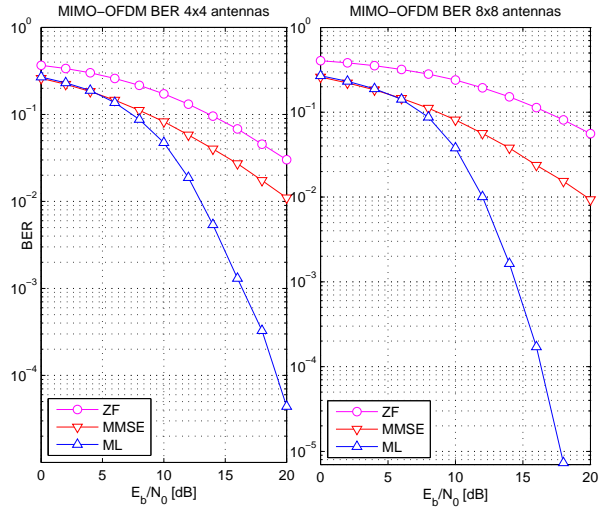}
                \vspace{-4mm}
                \caption{MIMO-OFDM comparison equipped with the ML, ZF and MMSE detectors, 4-QAM modulation and a) $4 \times 4$; b) $8\times 8$ antennas.}
                \label{fig:mimo_ofdm_compare}
                \end{figure}

                Fig. \ref{fig:mimo_ofdm_compare} compares the linear ZF against MMSE performance and how far such linear OFDM-MIMO detector performances are from the optimum ML detector, considering a typical $4\times4$ antennas and 4-QAM. One can conclude that for a medium-high SNR regime, MMSE detector provide a 3dB gain regarding ZF equaliser. Moreover, increasing the number of antennas, Fig. \ref{fig:mimo_ofdm_compare}.b),  the ML detector with $8 \times 8$ antennas provided much better performance than ML $4\times 4$ antennas but at cost of a huge increasing of complexity. ML detector makes an exhaustive search and chooses the best combination. However, the search space increases exponentially with the problem dimension; hence, for $8\times 8$, the search space was many times greater than $4\times 4$ antennas, which increasing substantially the simulation time and computational resources.

\subsubsection{{Input Parameter Calibration} for PSO-aided MIMO-OFDM Detector}\label{subsec:inputCalibPSO}
\begin{figure}[!htb]
\centering
\includegraphics[width=.502\textwidth]{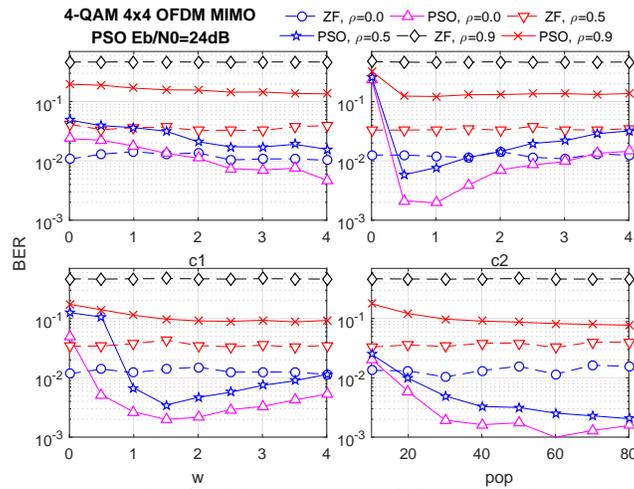}
 \vspace{-7mm}
\caption{Calibration of PSO input parameters values for 4-QAM $4 \times 4$ MIMO-OFDM detection problem operating under medium-high SNR and different spatial correlation indexes.} 
\label{fig:pso_calib_4x4}
\end{figure}

First, a round-trip of simulations were executed to tune PSO input parameters. Herein, these parameters are obtained numerically, in 100 simulation runs and taking the average values to obtain Fig. \ref{fig:pso_calib_4x4}.  Considering {start} parameters of $N_{\mathrm{pop}} = {40}; c_1 = c_2 = {2;w = 1}; N_{\mathrm{iter}} = 50$. In Fig. \ref{fig:pso_calib_4x4}, PSO input parameters were altered considering a wide range of input parameter values. The scenario assumed was $4\times 4$, $4-$QAM modulation MIMO-OFDM, considering the system operating in a medium-high SNR, {\it i.e.}, $E_b/N_0 = 24$dB and different values of spatial correlation. Choosing PSO parameters that provide small values of BER have resulted in the input parameters shown in Table \ref{tab:mimo_OFDM} and deployed in the numerical simulations setup discussed along this section. {Related to the population size, even with marginal decrease in BER, low values of $N_{\rm pop}$ are desired, since it has a direct impact in the computational complexity of the algorithm, as detailed in subsection \ref{ref:complexity}. }

\begin{figure}[!htb]
\centering
\includegraphics[width=0.49\textwidth]{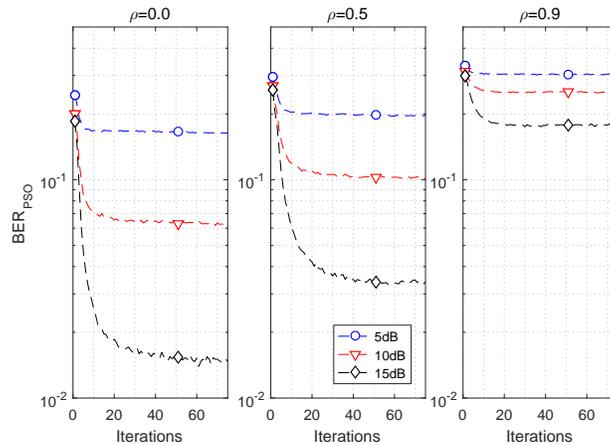}
\vspace{-7mm}
\caption{Convergence analysis for 4-QAM, $4 \times 4$ MIMO-OFDM with PSO detector considering different values of $E_b/N_0$. }
\label{fig:pso_convergence_4x4}
\end{figure}

In Fig. \ref{fig:pso_convergence_4x4}, the convergence {behavior} for the PSO-based detector is {analyzed}. It can be observed that convergence depends on $E_b/N_0$ level;  the  number of iterations for convergence increases with SNR, from {$\approx 25$} to {$50$} iterations when $E_b/N_0$ increases from 5dB to  10dB and 15dB. Moreover, {high values of} spatial correlation level {($\rho = 0.9$)} seem to interfere substantially in the convergence speed of the PSO algorithm applied in the MIMO-OFDM detection problem. After around {40} iterations, there are small improvements in the solution (symbol detection) provided by PSO algorithm for any spatial correlation level.\\

\subsubsection{{Input Parameter Calibration}  DE-aided MIMO-OFDM Detector}\label{subsec:inputCalibDE}
\begin{figure}[!htb]
    \centering 
    \includegraphics[width=.49\textwidth]{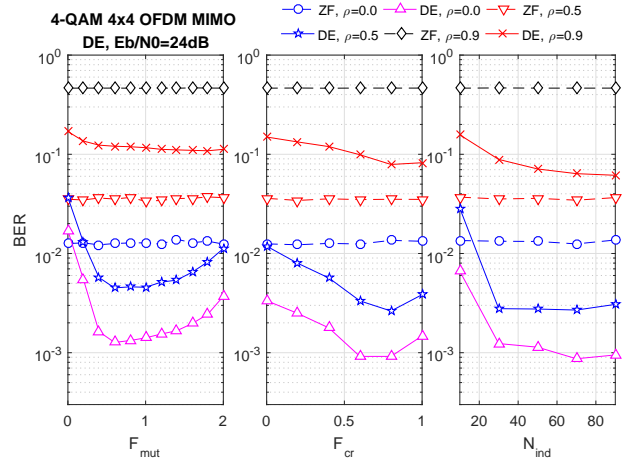}
    \vspace{-5mm}
    \caption{Calibration of input parameters for the DE-aided MIMO-OFDM detector algorithm considering different values of spatial correlation.}
    \label{fig:deVarParRound1-2}
 \end{figure}

A similar procedure is {carried out to find the best input parameter values of the DE-based detector in the context of MIMO-OFDM detection}. {The algorithm requires the parameters to be inside the interval} $F_{\rm cr} \in [0, 1]$ and $F_{\rm mul} \in [0, 2]$; $N_{\rm ind} \geq 4$ and it is recommended\cite{Storn1997} that $N_{\rm ind}$ in between  $5N_{\rm dim}$ and $10N_{\rm dim}$.  The selected {input parameters values were chosen as such those minimize the BER, and presented in Table \ref{tab:mimo_OFDM}. Notice that the optimum mutation factor value {increases} with the antenna correlation index $\rho$}. Fig.  \ref{fig:deVarParRound1-2} depicts the simulated BER curves for a wide range of input parameter values {showing the best values of such input parameters, {\it i.e.}, those values that minimizes the BER. The calibration procedure is finished when the range of those input parameters is narrowed.}

\begin{figure}[!htb]
   \centering 
   \includegraphics[width=.5\textwidth]{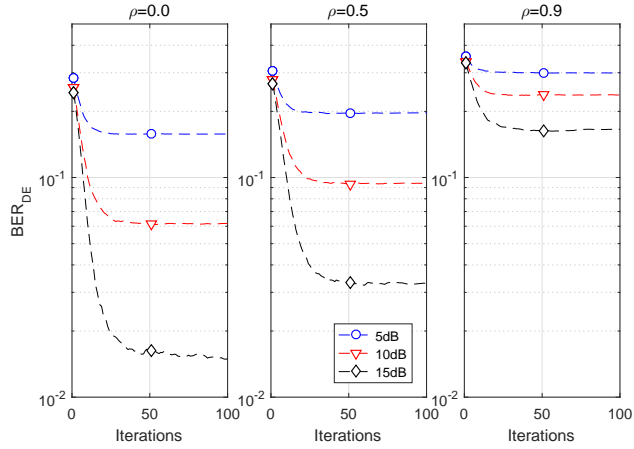}
   \vspace{-5mm}
   \caption{{Convergence of DE-aided detector for MIMO-OFDM systems for different spatial correlation values.} }
   \label{fig:de_convergence4x4_4QAM}
\end{figure}

{After input parameters tuning procedure, the convergence of the DE-aided OFDM-MIMO detector algorithm is obtained, as depicted in Fig.\ref{fig:de_convergence4x4_4QAM}. Similar to the PSO converge behavior, the convergence of the DE detector seems to be attained around 40 iterations, being influenced mainly by the $E_b/N_0$ levels.}

\subsubsection{Spatial Correlation Impact on the Performance}
\begin{figure}[!htb]
\centering
\hspace{-3mm}\includegraphics[width=.504\textwidth]{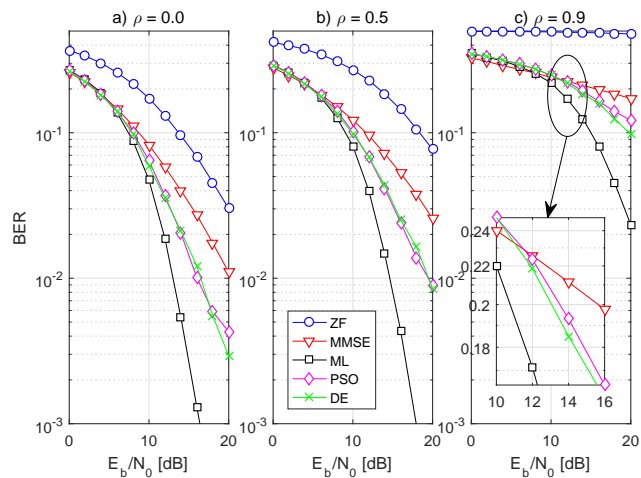}
\vspace{-5mm}
\caption{{BER performance for 4-QAM, $4 \times 4$ linear array (ULA) antennas MIMO-OFDM detectors under different values of spatial correlation and SNR.}} 
\label{fig:mimo_ofdm_correlation}
\end{figure}

In this subsection, the numerical simulation results for the BER performance were obtained under different correlation indexes $\rho$ values, representing the antennas separation in the transmitter and receiver side as depicted in Fig. \ref{fig:mimo_ofdm_correlation}. As inferred previously from  Figs. \ref{fig:pso_convergence_4x4} and \ref{fig:pso_calib_4x4}, the spatial correlation deteriorates the BER performance; as $\rho$ increases, the probability of error also increases. Under highest correlated channels $\rho = 0.9$, ZF detector provides a unacceptable performance, even operating under high $E_B/N_0$ region. The degrading impact of spatial correlation on the performance also influences the ML detector performance; however, the ML detector still attains a suitable performance considering uncoded system, at cost of an enormous computational complexity.  Alternatively, considering low-complexity evolutionary DE-based and PSO-aided detectors under $\rho=0$ scenario, PSO can outperform the MMSE; however, in the highly correlated situation, such performance advantage becomes marginal, {while DE-based MIMO-OFDM detector performs marginally worst than MMSE for all SNR regions.} Hence, under medium or even high correlated MIMO channels, the linear MMSE and the PSO-based detectors represent good options regarding performance-complexity tradeoff {in MIMO-OFDM systems}. 

\begin{figure}[!htb]
\centering
\includegraphics[width=.5\textwidth]{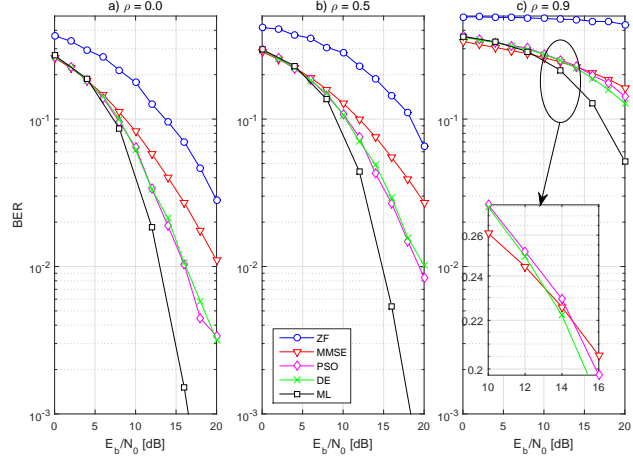}
 \vspace{-6mm}
\caption{{BER performance for 4-QAM $4 \times 4$  OFDM MIMO with linear and heuristic detectors considering planar array (URA) configuration and different values of correlation and SNR.}  } 
\label{fig:planar_performance_4x4}
\end{figure}

Fig. \ref{fig:planar_performance_4x4} explores  the BER performance considering {planar arrays (URA, instead of linear array)}. For high $E_b/N_0$, medium $rho$ and a low number of antennas (4 $\times $ 4), the planar array configuration slightly outperforms the linear array design for ZF, MMSE and PSO detectors (compare the BER performance of Fig. \ref{fig:mimo_ofdm_correlation} and Fig. \ref{fig:planar_performance_4x4}). Note that the use of the URA system implies a slightly higher correlation among antennas compared to the ULA. Despite this, the URA performance remains very similar to the ULA and is even slightly better at high SNR. 

\subsubsection{Sensibility Analysis}
{To compare the BER degradation w.r.t. array antenna correlation, the sensibility of the detectors's performance regarding the level of correlation can be defined as:}
\begin{equation}
\kappa_{\rm scn} = \log_{10}\textsc{ber}_{\rm scn} - \log_{10}\textsc{ber}_{\rm ref},
\end{equation}
where $\textsc{ber}_{\rm ref}$ represents the reference BER value, and $\textsc{ber}_{\rm scn}$ the BER at a specific scenario, including spatial correlation condition or detector type. 

\begin{figure}[!htb]
\centering
\includegraphics[width=.5\textwidth]{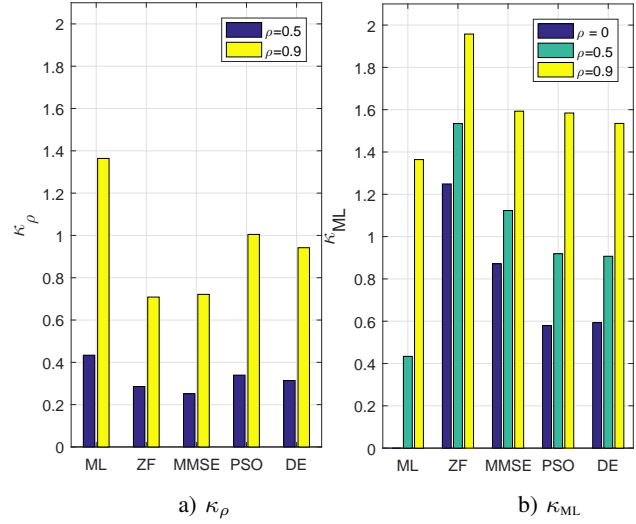}\\
\small \hspace{15mm} a) $\kappa_{\rho}$ \hspace{36mm} b) $\kappa_{\textsc{ml}}$
 \vspace{-2mm}
\caption{{Sensibility of detectors for two scenarios of correlation: a) $\kappa_{\rho}$, comparing each detector with its BER under $\rho = 0$. b) $\kappa_{\textsc{ml}}$, comparing with ML detector performance under $\rho = 0$.} } 
\label{fig:sensibility}
\end{figure}

For illustration purpose, two cases are studied: the degradation in performance comparing the BER of each detector w.r.t. uncorrelated antennas ($\rho=0$); and the degradation using the ML detector as reference, since its performance is superior to the others. Fig. \ref{fig:sensibility} depicts both sensibility scenarios:

\noindent\underline{$\kappa_{\rho}$}: In Fig. \ref{fig:sensibility} a), the sensibility considering the performance of each detector at $\rho = 0$ as the correlation increases is numerically obtained. Hence, comparing the performance degradation sensibility for each detector at $\rho = 0.5$ and $\rho=0.9$, one can conclude that the ML sensibility to the channel correlation increasing is severely degraded compared with the linear and heuristic detectors due to it excellent performance under $\rho = 0$ condition;  while for ZF detector, the degradation is small, since it already has poor performance compare to other detectors. In shorting, the four MIMO-OFDM detectors presents no-robustness to  the spatial correlation channel effect. 

\noindent\underline{$\kappa_\textsc{ml}$}:  In Fig. \ref{fig:sensibility} b), sensibility taking as reference the ML detector BER performance with $\rho = 0$ as $\textsc{ber}_{\rm ref}$. For medium correlation values ($\rho = 0,5$), the PSO was most near to ML sensibility performance degradation, and so $\kappa_\textsc{ml}$ has resulted relatively low. For $\rho = 0.9$, ZF detector performs poorly in terms of BER, resulting in high value of sensibility index $\kappa_\textsc{ml}$. PSO-aided detector is more sensible in terms of $\kappa_{\rho}$, since its BER vary more as correlation increases, but less sensible in terms of $\kappa_\textsc{ml}$, mainly for low and medium spatial correlation channels indexes ($\rho\leq 0.5$).

\subsection{{Complexity Analysis}}\label{ref:complexity}
{To evaluate the complexity of the algorithms, the number of \textit{floating points operation} (\textsc{flop}), defined as a floating point add, subtract, multiply or divide \cite{golub2012} between real numbers have been considered. Herein, Hermitian operator and \texttt{if} conditional operator were disregarded. In a real implementation, some platforms may use hardware random number generators, where an electric circuit is responsible to provide the random numbers; hence, the \textsc{flop} cost for random numbers generation was also disregarded in this analysis.}

\begin{figure}[!htb]
\centering
\includegraphics[width=0.49\textwidth]{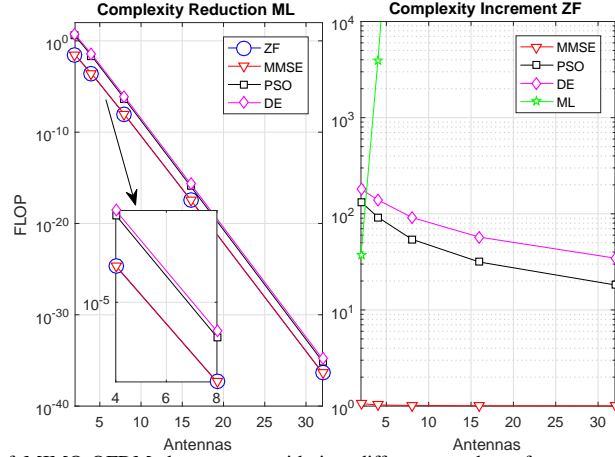}
\vspace{-7mm}
\caption{{Relative complexity of MIMO-OFDM detectors considering different number of antennas for linear and heuristic detectors in point-to-point scenario $N_t = N_r, N_{\rm dim} = 2N_t, N_{\rm pop} = N_{\rm ind} = 5\cdot N_{\rm dim}, \mathcal{I} = 50, \mathcal{M} = 4$.} }
\label{fig:flops}
\end{figure}

{The number of \textsc{flop}s required for the main operations is summarized in Table \ref{tab:referenceFlops} and the full complexity expressions denoted by $\Upsilon$, for the considered MIMO-OFDM detectors are presented in Table \ref{tab:flopLinearHeuristicDetectors}. To analyse the detector \textsc{flop}s complexity for a different number of antennas, Fig.\ref{fig:flops} depicts the linear and heuristic detector complexities assuming $N_{\dim} = 2N_t; N_t = N_r; N_{\rm ind} = N_{\rm pop} = 5\cdot N_{\dim}$ and assuming the number of iterations till the convergence obtained through simulations shown in Fig. \ref{fig:pso_convergence_4x4} and \ref{fig:de_convergence4x4_4QAM}.}

\begin{table}[htbp]
  \centering
  \caption{{Number of \textsc{flop}s for vector and matrices operations: ${\bf w}\in \mathbb{R}^{q \times 1}, {\bf A}\in\mathbb{R}^{m\times q}, {\bf B}\in\mathbb{R}^{q\times p}, {\bf C}\in\mathbb{R}^{m\times p}, {\bf D}\in\mathbb{R}^{q\times q}$. } }
  \label{tab:referenceFlops}
  {\renewcommand\arraystretch{1.4}
    \begin{tabular}{p{0.34\textwidth}p{0.1\textwidth}}
    \toprule
    {\bf Operation} & {\# \textsc{flop}s}\\
    \midrule
      {Matrix-matrix  multiply ${\bf AB}$} & {$mp(2q-1)$} \\
    {Matrix-vector multiply {\bf Aw}} & {$m(2q-1)$} \\
    {Matrix multiply-add ${\bf AB+C}$} & {$2mpq$} \\
    {Square root $\sqrt{.}$}  & {8} \\
    {Matrix inversion with LU factorization of {\bf D} \cite{BoydNumerical}} & {$2/3q^3 + 2q^2$} \\
    {Norm-2, $\sqrt{{\bf w}^T{\bf w}}$}	& {$2q-1 + 8$}\\
    \bottomrule
    \end{tabular}
    }
\end{table}%

The ML detector computes all possible input matrices \cite{Goldsmith2005} resulting in the evaluation of eq. \eqref{eq:MLD} $\mathcal{M}^{2N_t\times 1}$ times, where $\mathcal{M}$ representing the modulation order, resulting in the most computationally complex among the detectors considered. It can be observed that DE algorithm requires more \textsc{flop}s than PSO since it evaluates $2N_{\rm pop}$ times the fitness function per iteration in eq. \eqref{eq:selection} for individuals and crossover vectors. The complexity between the linear detectors are almost the same, differing from each other by an scalar-matrix multiplication and matrix-matrix sum in eq. \eqref{eq: ZFD} and eq. \eqref{eq:MMSED}. 

\begin{table}[htbp]
  \centering
  \caption{ {Number of \textsc{flop}s per subcarrier for the MIMO-OFDM detectors, with $\boldsymbol{\mathcal{H}} \in \mathbb{R}^{2N_r \times 2N_t}$, $\textswab{y} \in \mathbb{R}^{2N_r \times 1}$, $N_{\rm dim} = 2N_t$.} }
  \label{tab:flopLinearHeuristicDetectors}
  {\renewcommand\arraystretch{1.4}
    \begin{tabular}{p{0.15\textwidth}p{0.28\textwidth}} 
    \toprule
  {\bf Detector} & {\bf Number of Operations} \\
    \midrule
    {$\Upsilon_\textsc{ZF}(N_t, N_r)$}     & {$\dfrac{16}{3}N_t^3 + 4N_t^2 + 32N_t^2 N_r + 4N_tN_r - 2N_t$ } \\
    {$\Upsilon_\textsc{MMSE}(N_t, N_r)$}     & {$\dfrac{16}{3}N_t^3 + 8N_t^2 + 32N_t^2N_r + 4N_tN_r$ } \\
    {$\Upsilon_\textsc{PSO}(N_t, N_r, N_{\rm pop}, \mathcal{I})$}     & {$N_{\rm pop} \mathcal{I} ( 8N_tN_r + 20N_t + 4N_r + 7)$ } \\
    {$\Upsilon_\textsc{DE}(N_t, N_r, N_{\rm ind}, \mathcal{I})$}     &  { $N_{\rm ind} \mathcal{I} (16N_tN_r + 12N_t + 8N_r + 14) $}\\
    {$\Upsilon_\textsc{ML}(N_t, N_r, \mathcal{M})$}     &  { $\mathcal{M}^{2N_t} ( 8N_tN_r + 4N_r + 7) $}\\
        \bottomrule
 {$\mathcal{M}:$ modulation order} & {$\mathcal{I}:$ \# iterations heuristic algorithms}\\
\end{tabular}
}
\end{table}

{A plot of relative complexity is depicted in Fig.\ref{fig:flops}. On the left side, the complexity reduction  relative to ML and linear/heuristic detectors evaluated as $\Upsilon_{\textsc{det}}/\Upsilon_{\textsc{ml}}$ is shown. All the studied MIMO-OFDM detectors provide a complexity decreasing regarding the ML detector. Note that PSO slightly provides more reduction than DE, and linear detectors more than heuristics, at the cost of BER performance. On the right side, the complexity increasing relative to linear low-complexity ZF MIMO-OFDM detector $\Upsilon_{\textsc{det}}/\Upsilon_{\textsc{zf}}$ is determined. Worth to noting that the linear MMSE detector has complexity near to the ZF resulting in values next to 1; while ML detector increases rapidly with the increase in the number of antennas. The heuristic PSO detector provides lower increments in complexity than DE detector and almost the same BER performance, offering a good complexity tradeoff between computational complexity versus performance, mainly when the number of antennas increases (massive MIMO systems). 
}

\section{Conclusions}\label{sec:conclusions}
Analysis of an OFDM scheme is developed considering NLOS Rayleigh fading channel conditions; for that, we have implemented and validated uncorrelated multipath fading generation based on TD Jakes modified model together with a generic power-delay profile. JM model is able to generate suitable uncorrelated delayed fading channel coefficients thanks to the WH matrices introduced in the modified version of the model. As the number of subcarriers increases, the performance increases too until the sub-channel flatness condition is achieved. The cost of this is a longer computational time and complexity requirement. In terms of CP size, one should take into account the ISI effect, which must to be combated, but there is a cost to do this (waste of energy). 

In MIMO-OFDM systems with multiplexing configuration, ZF detector is less complex than MMSE and, as expected, provides worse performance, specially in spatial correlated channels. For instance, in correlated channels with $\rho = 0.9$, ZF BER performance is very poor; the probability of error do not improve significantly with the increase of SNR. 
Changing the number of antennas, while assuming perfect knowledge of the channel, the MMSE detector is able to provide similar BER performance for the three configurations in terms of spatial correlation. Moreover, note that the performance of SISO-OFDM is better than the performance of MIMO-OFDM. This is due to the fact that the MIMO-OFDM system with linear detection operates in multiplexing mode is not able to offer diversity gain. On the other hand, in multiplexing mode the system is able to provide a linear increment in the capacity regarding the number of antennas.

Extensive simulations were deployed and suitable evolutionary heuristic PSO {and DE} input parameters were chosen numerically for the MIMO-OFDM detection problem. The convergence of PSO-based  detector depends {mainly} on $E_b/N_0$ level, requiring more iterations as SNR increases.

The spatial correlation degrades the performance of the {analyzed} MIMO-OFDM detectors. For uncorrelated scenario ($\rho = 0$), PSO-aided detector outperforms linear detectors ZF and MMSE. However, for high correlation ($\rho=0.9$), PSO detector gain in terms of BER performance becomes marginal. The performance degradation as correlation increases is quantified by the sensibility of the detectors for different levels of correlation.

Planar antenna arrays marginally outperform the linear array configurations for ZF, MMSE, PSO { and DE} MIMO-OFDM detectors considering  high SNR operation region, and low number of antennas. When the number of antennas increases, such outperforming {may} becomes noticeable. Although the correlation among antennas is slightly higher in the URA, this difference is not enough to deteriorate the performance of the system.

{Comparing the complexity of the detector algorithms, the linear MMSE detector provides better performance than the linear ZF with almost same computational complexity. Among the representative evolutionary heuristic MIMO-OFDM detectors, the PSO provides lower increments in complexity regarding the DE detector, and almost the same (similar) BER performance for all system and channel scenarios analysed, both offering a suitable computational complexity {\it versus} performance tradeoff, even under medium spatial antenna correlation levels.}

\section*{Acknowledgment}
This work was supported in part by the National Council for Scientific and Technological Develop- ment (CNPq) of Brazil under Grants 130464/2015-5 (Scholarship) and 304066/2015-0 (Researcher grant), in part by Araucaria Foundation, PR, under Grant 302/2012 (Research) and by Londrina State University - Paraná State Government, Brazil.


\end{document}